\documentclass[twocolumn,showpacs,preprintnumbers,amsmath,amssymb]{revtex4}

\def\undersim#1{\setbox9\hbox{${#1}$}{#1}\kern-\wd9\lower
    2.5pt \hbox{\lower\dp9\hbox to \wd9{\hss $_\sim$\hss}}}
\usepackage{amssymb}
\usepackage{amsmath}
\usepackage{graphicx}
\usepackage{color}

\def\undersim#1{\setbox9\hbox{${#1}$}{#1}\kern-\wd9\lower
    2.5pt \hbox{\lower\dp9\hbox to \wd9{\hss $_\sim$\hss}}}

\def\mv{{\mathbf v}}
\def\mr{{\mathbf r}}

\def\mr{{\mathbf r}}

\def\mk{{\mathbf p}}

\begin{document}

\title{Squeezed spectra and back-to-back correlations of protons and antiprotons at RHIC energies}

\author{Yong Zhang}
\affiliation{\footnotesize School of Mathematics and Physics, Jiangsu University of Technology, Changzhou, Jiangsu 213001, China\\}

\begin{abstract}
This study constrains the range of in-medium mass modification through a comparison of theoretical calculations with experimental transverse momentum spectra and the yield ratio $\bar{p}/p$ of protons and antiprotons. Based on the constrained range and a Gaussian source model with radial flow, the theoretical predictions for the fermion back-to-back correlation (fBBC) of $p\bar{p}$ pairs at RHIC energies are presented. The results reveal a strong sensitivity of the fBBC signal to the assumed source time distribution: a Lorentzian form generates a pronounced high-momentum signal, whereas an $\alpha$-stable L\'{e}vy form leads to a marked low-momentum signal. Moreover, the in-medium mass modification is shown to enhance the yield ratio $\bar{p}/p$. Therefore, events characterized by a larger $\bar{p}/p$ ratio are predicted to have a significantly higher probability of exhibiting a detectable fBBC signal.
This study may propose a promising new direction for the experimental observation of this phenomenon.

Keywords: Protons and antiprotons; in-medium mass modification; spectra; fermion back-to-back correlation.

\end{abstract}

\pacs{25.75.-q, 21.65.jk, 25.75.Dw, 25.75.Gz.}
\maketitle

\section{Introduction}
In relativistic heavy-ion collisions, where a hot, dense medium is expected to form,
the in-medium properties of particles have been a topic of enduring interest.
{\color{black}Recently, the PHENIX Collaboration reported a measurement of the centrality dependence of L\'{e}vy-stable two-pion Bose-Einstein correlations in $\sqrt{s_{NN}}$ = 200 GeV Au+Au collisions \cite{prc2024}. The study found that the data are not inconsistent with a significant in-medium mass reduction of the ${\eta}'$ meson by nearly 400 MeV,
as a consequence of a special chiral symmetry restoration in a hot and dense hadronic medium \cite{tcs1998,tcs2010,tcs2011}. Although the evidence is indirect, it provides a valuable motivation:
if the ${\eta}'$ meson can undergo such a significant mass modification, other hadrons may also experience modifications. Thus, it is necessary to
search for other indirect signatures of in-medium mass modifications.}
Particle interactions with the surrounding medium can generate a squeezing effect,
resulting in a squeezed back-to-back correlation between particles and their antiparticles \cite{AsaCsoGyu99,Padula06,Padula10,Padula10a,Zhang15a,YZHANG_CPC15,XuZhang19,fm-PLB,fm-NPA}.
This squeezing effect is linked to in-medium mass modifications of fermions (or bosons) via the Bogoliubov-Valatin (or Bogoliubov) transformation, which establishes a direct relationship between the creation (annihilation) operators of particles in the medium and their vacuum counterparts \cite{AsaCsoGyu99,Padula06,Padula10,Padula10a,Zhang15a,YZHANG_CPC15,XuZhang19,fm-PLB,fm-NPA}.
Investigating this squeezing effect and the squeezed back-to-back correlation could offer new insights into {\color{black}the in-medium mass modification of hadrons, as well as} the dynamic and thermal characteristics of the particle source generated in relativistic heavy-ion collisions.

The measurement of the squeezed fermion back-to-back correlation (fBBC) of $p\bar{p}$ was performed in $0-25\%$ Au+Au collisions at $\sqrt{s_{NN}}$ = 200 GeV
within the transverse momentum range of $0.6-0.9$ GeV/$c$. However, the measured fBBC strength is approximately $2\%$, which is not present on a significant level \cite{BBCsy}.
There are two possible reasons for this: either there is a mass modification of protons and antiprotons, but the suppression effect caused by the freeze-out distribution renders it practically unobservable, or there is no mass modification within the transverse momentum range of $0.6-0.9$ GeV/$c$ in $0-25\%$ Au+Au collisions at $\sqrt{s_{NN}}$ = 200 GeV \cite{BBCsy}. Whether the fBBC is observable in other collision systems or other momentum ranges remains to be established and requires further investigation.

The single-particle invariant momentum distribution is also affected by the squeezing effect arising from in-medium mass modifications \cite{AsaCsoGyu99,Padula06,fm-PLB,Zhy-ijmpe2025}.
Recent studies have shown that the squeezing effect reduces the slope of transverse momentum spectra in both protons and antiprotons.
Furthermore, under conditions of nonzero proton chemical potential, the effect manifests differently for each particle type and
leads to an increased yield for both, a phenomenon especially evident for antiprotons at higher transverse momentum.
Consequently, the squeezing effect results in the $\bar{p}/p$ yield ratio increasing with transverse momentum, eventually approaching unity \cite{Zhy-ijmpe2025}.
The differing impact of the squeezing effect on spectra and yield ratio under various degrees of in-medium mass modification enables the estimation of this modification through a comparison of theoretical and experimental results.
Since a momentum-independent in-medium mass modification cannot theoretically describe the experimental data on spectra and yield ratio under squeezing effects \cite{Zhy-ijmpe2025},
this paper adopts a momentum-dependent in-medium mass modification to further investigate the influence of the squeezing effect on these observables.
Based on a comparison of theoretical spectral data and yield ratio with experimental data, this paper determines the range of in-medium mass modification and provides a theoretical prediction of the fBBC of $p\bar{p}$ at the Relativistic Heavy Ion Collider (RHIC) energies.
To account for the inhomogeneous and expansive nature of systems created in relativistic heavy-ion collisions,
a Gaussian source with radial flow is employed to describe particle emission in the calculations.
This source model has been widely employed to study squeezed back-to-back correlations of boson-antiboson pairs in relativistic heavy-ion collisions \cite{Padula06,Padula10,Padula10a,YZHANG_CPC15}. Recently, it was further applied to investigate how expanding flow affects the fBBC of proton-antiproton and $\Lambda$-$\bar{\Lambda}$ pairs \cite{Zhy-universe}. Therefore, it may serve as a reasonable source model for this study.

{\color{black}The results presented in Fig. 3 and Table I show that the STAR experimental data on proton and antiproton single-particle spectra, as well as their ratios at six distinct collision energies at RHIC, are not inconsistent with a momentum-dependent in-medium mass modification affecting both protons and antiprotons. These results, combined with similarly indirect observations of the PHENIX experiment reported in Ref. \cite{prc2024}, point to the importance of searching for additional observables that can indirectly probe the in-medium mass modification of hadrons, as well as the related back-to-back correlations of stable or long-lived particles that do not decay significantly in the hot hadronic medium created in relativistic heavy-ion collisions.}

This paper's theoretical predictions indicate that the $p\bar{p}$ fBBC signature is highly sensitive to the source time distribution; a Lorentzian form \cite{AsaCsoGyu99,Padula06,Padula10,Padula10a,fm-PLB} generates a pronounced high-momentum signal, while an $\alpha$-stable L\'{e}vy form \cite{Padula10} results in a pronounced low-momentum signal.
An in-medium mass modification is predicted to enhance the $\bar{p}/p$ yield ratio. Consequently, events with a larger $\bar{p}/p$ within the same collision system are likely to exhibit a higher probability of fBBC detection. This study may offer a new avenue for its experimental observation.

The rest of this paper will proceed as follows.
Section II introduces the fBBC correlation function for fermion-antifermion pairs and provides the formulae for the single-particle invariant momentum distributions of fermions and antifermions from expanding sources.
In Section III, simulated spectral data are compared with experimental results, leading to an extracted value for the in-medium mass modification. Additionally, predictions for the fBBC strength of $p\bar{p}$ pairs in central Au+Au collisions at RHIC energies are presented. Finally, a summary and discussion are provided in Section IV.

\section{Formulas}
Let $a_\mk$ and $a^\dagger_\mk$ ($\bar{a}_\mk$ and $\bar{a}^\dagger_\mk$)
represent the annihilation and creation operators for free fermions (antifermions) with momentum $\mk$,
mass $m_0$, and energy $\omega_{\mk} = \sqrt{\mk^2 + m_0^2}$.
Correspondingly, let $b_\mk$ and $b^\dagger_\mk$ ($\bar{b}_\mk$ and $\bar{b}^\dagger_\mk$) denote the annihilation and creation operators of the fermions (antifermions) with momentum $\mk$, modified mass $m_{\!*}$, and energy $\Omega_\mk=\sqrt{\mk^2 + {m_*}^2}$ in the medium.
Here, $m_{\!*} \neq m_0$ indicates that the mass of the fermions (antifermions) has been modified due to their interaction with the medium.
A fermionic Bogoliubov-Valatin transformation connects the annihilation and creation operators in vacuum with those in a medium \cite{fm-PLB,fm-NPA,prc-bv}
\begin{equation}
\left(
\begin{array}{c}
a_{\lambda,{\mk}} \\
{\bar a}^{\dagger}_{\lambda^\prime,-{\mk}} \end{array}
\right) =
\left(
\begin{array}{cc}
c_{\mk} &
\frac{r_{\mk}}{|r_{\mk}|} \,s_{\mk} \,A  \\
-\frac{r^*_{\mk}}{|r_{\mk}|}\, s^*_{\mk}\, A^{\dagger} &
c^*_{\mk}
\end{array}
\right)
\left(
\begin{array}{c}
b_{\lambda,{\mk}} \\
{\bar b}^{\dagger}_{\lambda^\prime,-{\mk}}
\end{array}      \right ),
\label{BVtransf}
\end{equation}

\begin{equation}
c_{\mk}=\cos r_{\mk}, \,\,\,\,\,\,\,\,s_{\mk}=\sin r_{\mk},
\end{equation}

\begin{eqnarray}
\label{rp}
\tan(2 r_{\mk}) = - \frac{|{\mk}| \delta m }
{\omega^2_{\mk} - m_0 \delta m}.
\end{eqnarray}
The matrix elements $A_{\lambda,\lambda^\prime}$ of the $2~\times~2$ matrix $A$ are defined as $\chi^{\dagger}_{\lambda}\sigma\cdot{\hat{\mk}}\tilde
\chi^{\phantom\dagger}_{\lambda^\prime}$. Here, $\lambda$ and $\lambda^\prime$ represent the spin projections. The unit vector $\hat{\mk}$
is aligned with the direction of ${\mk}$, and $\chi$ is a Pauli spinor with $\tilde \chi= -i\sigma^2\chi$ \cite{fm-PLB}.
$\delta m$ represents the in-medium mass modification, and $\delta m$ = $m_0 - m_*$.
A non-zero $\delta m$ results in fBBC between fermion-antifermion pairs.

For an expanding and inhomogeneous system, the fBBC correlation function for fermion-antifermion pairs with back-to-back momenta $\mk$ and $-\mk$ can be expressed as \cite{fm-PLB,fm-NPA,AsaCsoGyu99,Padula06,Zhy-universe}
\begin{align}
\label{fbbcf1}
\hspace*{-1mm}C(\mk,-\mk)=1+F\Bigg|\!\int \frac{\rho(\mr)d^3r}{(2\pi)^3}\, \Bigl\{[1-n'(\mk')-\widetilde{n}'(\mk')]c_{\mk'}s_{\mk'}\Bigl\}\Bigg|^2\nonumber\\
\Bigl/\!\int \frac{\rho(\mr)d^3r}{(2\pi)^3}\, \! \Bigl\{|c'_{\mk'}|^2\,
n'(\mk')+\,|s'_{\mk'}|^2\,[\,1- \widetilde{n}'(\mk')]\Bigr\}\nonumber\\
\hspace*{8mm}\Bigl/\!\int \frac{\rho(\mr)d^3r}{(2\pi)^3}\, \! \Bigl\{|c'_{-\mk'}|^2\,
\widetilde{n}'(-\mk')+\,|s'_{-\mk'}|^2\,[\,1- n'(-\mk')]\Bigr\},\nonumber\\
\end{align}
\begin{equation}\label{csk}
c'_{\mk'}=\cos[\,r'_{\mk'}\,], \,\,\,s'_{\mk'}=\sin[\,r'_{\mk'}\,],
\end{equation}
\begin{eqnarray}
\label{rphyd}
\tan(2 r'_{\mk'}) = - \frac{|{\mk'}| \delta m }
{\omega'^2_{\mk'} - m_0 \delta m},
\end{eqnarray}
\begin{eqnarray}
&&\hspace*{-7mm}\omega'_{\mk'}=\sqrt{\mk'^2+m^2}=k^{\mu} u_{\mu}(r)\nonumber\\
&&\hspace*{0mm}=\gamma_\mv\,[\,\omega_{\mk}-\mk\cdot\mv(r)\,],
\end{eqnarray}
\begin{eqnarray}
\label{Omp}
&&\hspace*{-7mm}\Omega'_{\mk'}=\sqrt{\mk'^2+m_*^2}\nonumber\\
&&=\sqrt{[k^{\mu} u_{\mu}(r)]^2-m_0^2+m_*^2},
\end{eqnarray}
\begin{eqnarray}
\label{fmdhyd}
n'(\mk') = \frac{1}{\exp[(\Omega'_{\mk'}-\mu_i)/T+1]}, \\
\widetilde{n}'(-\mk') = \frac{1}{\exp[(\Omega'_{-\mk'}+\mu_i)/T+1]}.
\end{eqnarray}
Here, the four-velocity of the source at position $r$ is represented by $u_{\mu}(r)$.
The four-momentum of the fermion is given by $k^{\mu}=(\omega_{\mk},{\mk})$, where ${\mk}'$
denotes the local-frame momentum corresponding to $\mk$.
The fermion's in-medium chemical potential is denoted by $\mu_i$, and its freeze-out temperature by $T$.
$F$ is the finite time suppression factor introduced to model a more gradual freeze-out.
In this paper, two types of time distributions are considered: one is a Lorentzian form given by
$F = [1+(\omega_{\mk}+\omega_{-\mk})^2\Delta t^2]^{-1}$ \cite{AsaCsoGyu99,Padula06,Padula10,Padula10a,fm-PLB}, and the other is an
$\alpha$-stable L\'{e}vy form expressed as $F = \exp{\{-[\Delta t(\omega_{\mk}+\omega_{-\mk})]^{\alpha}\}}$ \cite{Padula10}.
$\Delta t$ is a parameter representing the width of the time distribution.
The exponent $\alpha$ is defined as a parameter in the $\alpha$-stable L\'{e}vy form.
The spatial distribution of the fermion emitting source is described by
$\rho(\mr)$, which is treated as {\color{black}\cite{plb1998, cejp2004}}
\begin{equation}
\label{fr}
\rho(\mr)= e^{-\mr^2/(2R^2)},
\end{equation}
The parameter $R$ quantifies the spatial width of the source, while the radial expansion velocity is defined as {\color{black}\cite{cejp2004,npa1978}}
\begin{equation}
\begin{cases}
v(\mr)= v_0\mr /R  \,\,\,\,\,\,\,\,\,\, r<R,\\
v(\mr)= v_0              \,\,\,\,\,\,\,\,\,\,\,\,\,\,\,\,\,\,\,\,  r\geq R.
\end{cases}
\end{equation}
$v_0$ sets the maximum radial flow velocity.
For the source under consideration, the observed single fermion and antifermion spectra are \cite{Zhy-ijmpe2025}
\begin{eqnarray}\label{SPexp}
 N(\mk)=\!\xi\int \rho(\mr)d^3r\frac{g_i}{(2\pi)^3}\omega_{\mk}\, \! \Bigl\{|c'_{\mk'}|^2\,
n'(\mk')\nonumber\\
+\,|s'_{\mk'}|^2\,[\,1-\widetilde{n}'(\mk')]\Bigr\},
\end{eqnarray}
\begin{eqnarray}\label{SPexp1}
 \widetilde{N}(\mk)=\!\xi\int \rho(\mr)d^3r\frac{g_i}{(2\pi)^3}\omega_{\mk}\, \! \Bigl\{|c'_{\mk'}|^2\,
\widetilde{n}'(\mk')\nonumber\\
+\,|s'_{\mk'}|^2\,[\,1- n'(\mk')]\Bigr\}.
\end{eqnarray}
The symbol $g_i$ denotes the degeneracy factor for fermion species $i$. $ N(\mk)$ and $\widetilde{N}(\mk)$ correspond to the observed spectra of the single fermion and single antifermion, respectively. The parameter $\xi$ controls the particle yield and is assumed identical for fermions and antifermions \cite{Zhy-ijmpe2025}.

\section{Results}
In this section, the influence of the squeezing effect on both the transverse momentum spectra of proton (\,$p$\,) and antiproton (\,$\bar{p}$\,) and their yield ratio $\bar{p}/p$ is first presented.
The in-medium mass modification, taken to be the same for both $p$ and $\bar{p}$, is treated as momentum-dependent and is represented by the parameter $\delta m(\mk)$ (\,$\delta m(\mk) = m_0 - m_* = \delta m_0 \exp[-\mk^2/{\Lambda}^2]$\,).
The parameter ${\Lambda}$ governs the momentum dependence of the in-medium mass modification, while $\delta m_0$ denotes the magnitude of this modification at zero momentum (\,$\mk = 0$\,) for both $p$ and $\bar{p}$.
The strength of the fBBC of $p\bar{p}$ is quantified by $\lambda_{BB} = C(\mk,-\mk)-1$ \cite{BBCsy}.
A value of $\lambda_{BB} = 0$ indicates the absence of a correlation from in-medium mass modification, while $\lambda_{BB} > 0$ signifies the presence of such a fBBC.
The strength of the fBBC ($\lambda_{BB}$) for $p\bar{p}$ with the momentum-dependent in-medium mass modification is also shown in this section.
The mass of the proton and antiproton in a vacuum $m_0$ and their freeze-out temperature $T$ are taken as 938.27 MeV \cite{PDG24} and 140 MeV \cite{fm-PLB}, respectively.

\begin{figure*}[htbp]
\hspace*{-15mm}\includegraphics[scale=0.7]{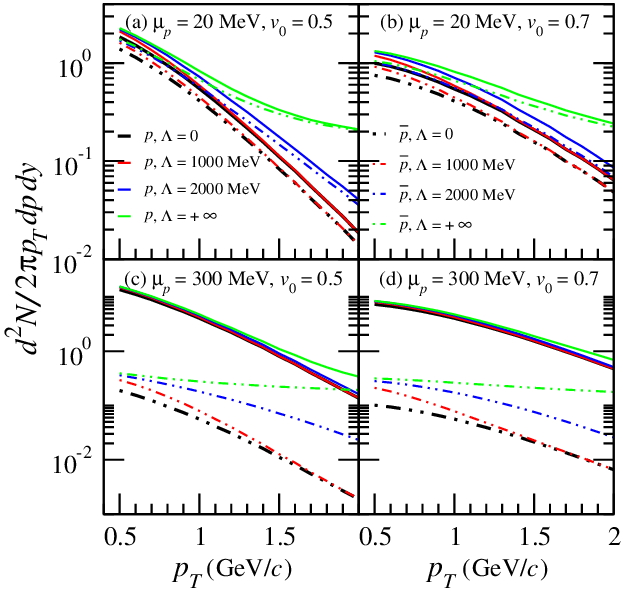}\hspace*{9mm}\includegraphics[scale=0.7]{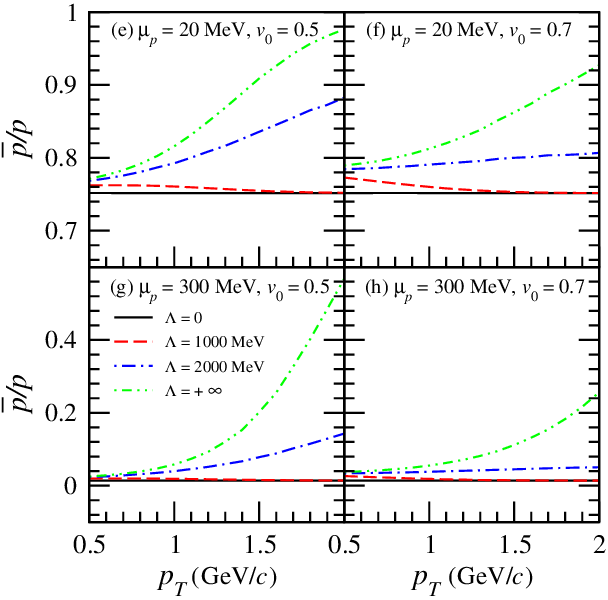}
\vspace*{-1mm}
\caption{(Color online) The transverse momentum spectra of protons and antiprotons for various $\Lambda$ values are shown on the left panel, and the yield ratios $\bar{p}/p$ are shown on the right panel.}
\label{pryx}
\end{figure*}

In Fig. \ref{pryx}, the transverse momentum spectra of protons and antiprotons for various $\Lambda$ values are shown on the left panel, and the yield ratios $\bar{p}/p$ are shown on the right panel. Here, $\delta m_0 = 20 $ MeV.
The case ${\Lambda} = 0$ gives $\delta m(\mk) = 0$ for any momentum (without squeezing effect), while ${\Lambda} = +\infty$ gives the constant result $\Delta m(\mk) = \delta m_0$.
In Figs. \ref{pryx}(a), (b), (e) and (f), the in-medium chemical potential of proton ($\mu_p$) is taken as 20 MeV, while in Figs. \ref{pryx}(c), (d), (g) and (h), it is set to 300 MeV. These values correspond approximately to the chemical potentials expected at collision energies of several GeV and 200 GeV, respectively.
For ${\Lambda} = 0$, which corresponds to the case without squeezing effect, the transverse momentum spectra of protons and antiprotons differ by a momentum-independent amount (see black lines in Figs. \ref{pryx}(a)-(d)). And the yield ratios $\bar{p}/p$ also remain constant as a function of momentum (see black lines in Figs. \ref{pryx}(e)-(h)).
For ${\Lambda} = 2000 $ MeV or ${\Lambda} = +\infty$, the squeezing effect increases the yields of protons and antiprotons, an enhancement particularly pronounced for antiprotons at high transverse momentum under a high chemical potential. Thereby, it reduces the difference between their spectra as transverse momentum increases (see blue and green lines in Figs. \ref{pryx}(a)-(d)). Furthermore,
the yield ratio $\bar{p}/p$ increases as the transverse momentum increases (see blue and green lines in Figs. \ref{pryx}(e)-(h)).
As the expansion velocity decreases, this phenomenon becomes more pronounced. Likewise, as the parameter ${\Lambda}$ increases, the squeezing effect has a more obvious impact on the transverse momentum spectra of protons and antiprotons. For ${\Lambda} = 1000 $ MeV, the squeezing effect only affects the yield ratios $\bar{p}/p$ in small transverse momentum regions.

\begin{figure*}[htbp]
\includegraphics[scale=0.69]{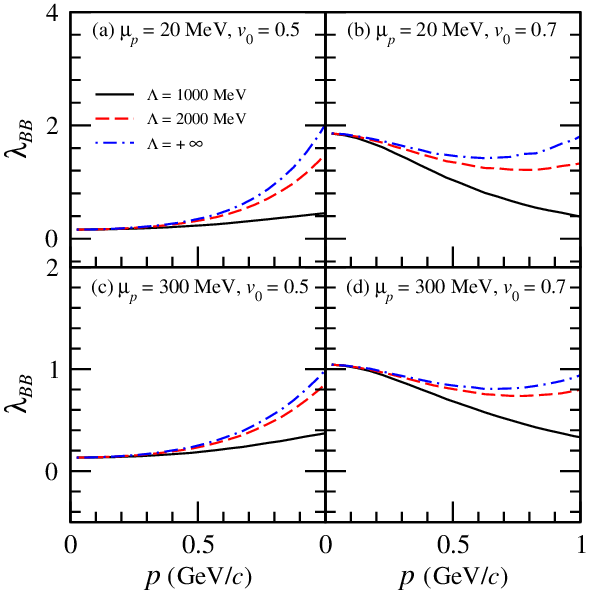}\hspace*{9mm}\includegraphics[scale=0.69]{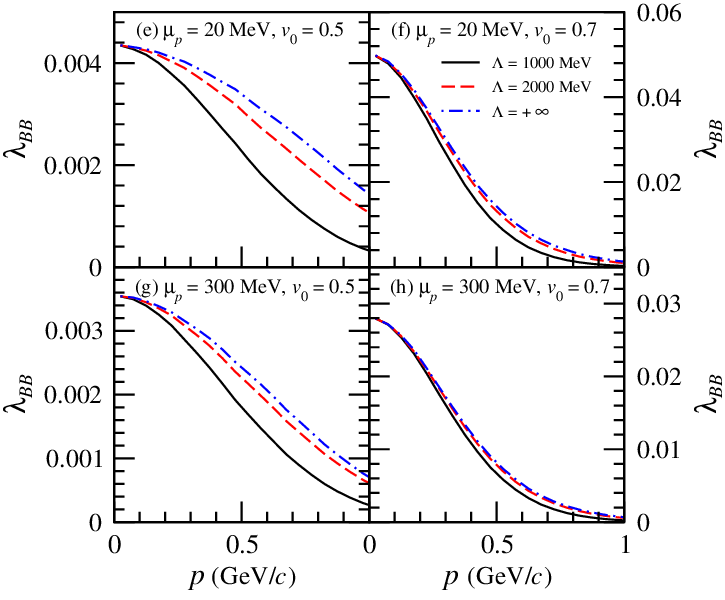}
\vspace*{-3mm}
\caption{(Color online) The left panel shows the fBBC strengths ($\lambda_{BB}$) of $p\bar{p}$ with a Lorentzian time distribution for various $\Lambda$ values, while the right panel shows those with an $\alpha$-stable L\'{e}vy time distribution.}
\label{ckyx}
\end{figure*}

In Fig. \ref{ckyx}, the fBBC strengths ($\lambda_{BB}$) for $p\bar{p}$ with Lorentzian and $\alpha$-stable L\'{e}vy form time distributions are shown in the left and right panels, respectively. Here, $\delta m_0 = 20 $ MeV. For a Lorentzian time distribution ($F = [1+(\omega_{\mk}+\omega_{-\mk})^2\Delta t^2]^{-1}$), $\Delta t$ is taken as 2 fm/$c$ \cite{fm-PLB}.
For sources with a Lorentzian time distribution, the behavior of the fBBC strengths varies with the velocity parameter $v_0$.
When $v_0$ = 0.5, the strengths increase with momentum. In contrast, for $v_0$ = 0.7 and $\Lambda$ = 1000 MeV, the strengths decrease as momentum increases, whereas for
$\Lambda \geq$ 2000 MeV, they first decrease and then increase with rising momentum. This phenomenon is attributed to the combined effects of source velocity \cite{Zhy-universe} and momentum-dependent mass modification on the fBBC strengths ($\lambda_{BB}$). For an $\alpha$-stable L\'{e}vy time distribution ($F = \exp{\{-[\Delta t(\omega_{\mk}+\omega_{-\mk})]^{\alpha}\}}$), the parameters are set as follows: $\Delta t = 1$ fm/$c$ and $\alpha = 1$ \cite{Padula10}.
For sources with an $\alpha$-stable L\'{e}vy time distribution, the fBBC strengths decrease as momentum increases. This phenomenon, which is similar to the results for
$K^+K^-$ in Ref. \cite{Padula10}, is mainly attributable to the $\alpha$-stable L\'{e}vy time distribution of the source.
The fBBC signal is significantly weaker for sources with an $\alpha$-stable L\'{e}vy time distribution compared to those with a Lorentzian one.

\begin{figure*}[htbp]
\includegraphics[scale=0.69]{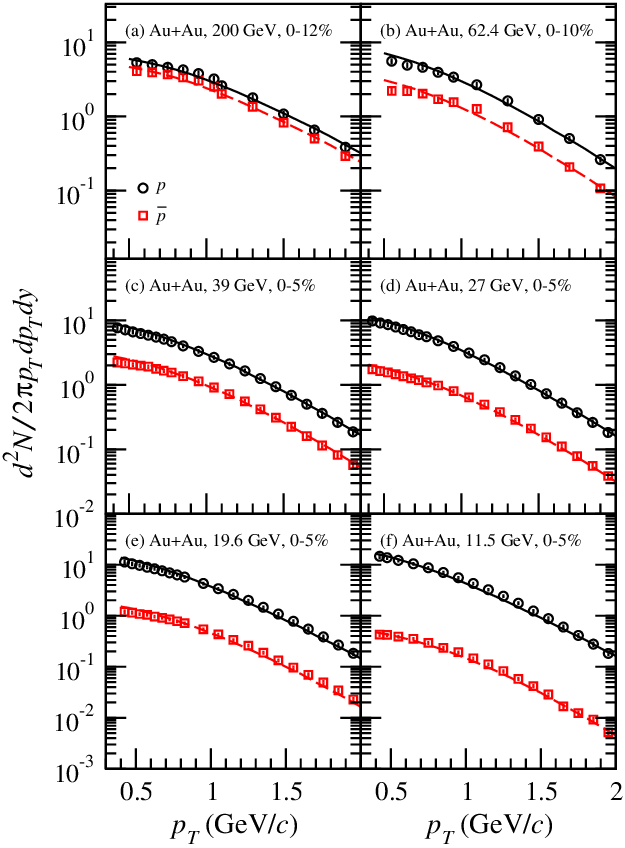}\hspace*{7mm}\includegraphics[scale=0.69]{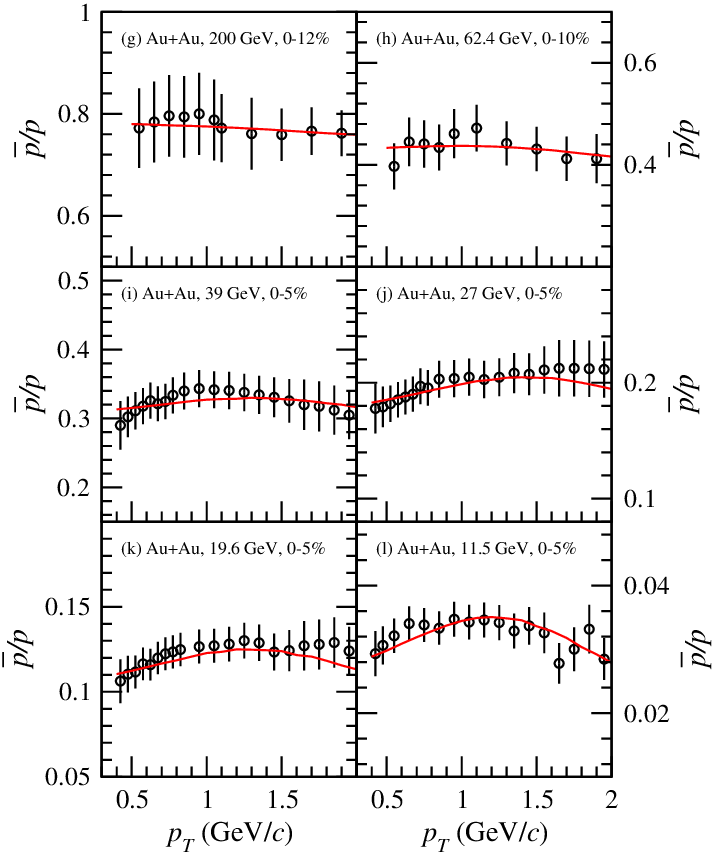}
\vspace*{-1mm}
\caption{(Color online) The left panel plots the transverse momentum spectra of protons and antiprotons for central Au+Au collisions at $\sqrt{s_{NN}}$ = 11.5, 19.6, 27, 39, 62.4, and 200 GeV, whereas the right panel depicts the corresponding yield ratios $\bar{p}/p$. Circle and square symbols denote the experimental data from the STAR Collaboration \cite{STAR_plbs,STAR_proton}, and lines show the simulated results.}
\label{ptr}
\end{figure*}

The mass modification of particles is not an observable quantity, existing only inside the hot and dense medium.
The fBBC strength is related to the in-medium mass modification. If the in-medium mass modification is zero, the fBBC strength will be zero.
To obtain a more reasonable estimate of the fBBC intensity, the simulated spectral data are compared with experimental data, deriving an in-medium mass modification value.
In Fig. \ref{ptr}, the transverse momentum spectra of protons and antiprotons in central Au+Au collisions at $\sqrt{s_{NN}}$ = 11.5, 19.6, 27, 39, 62.4, and 200 GeV are presented on the left panel, while the corresponding yield ratios $\bar{p}/p$ are shown on the right. The experimental data (circle and square symbols) are from the STAR Collaboration \cite{STAR_plbs,STAR_proton}. The lines represent results from the simulation. The simulation results closely match the experimental results,
with few exceptions in Figs. \ref{ptr}(b) and (l).
In the model used in this paper, proton and antiproton yields are governed primarily by the source radius $R$ and the parameter $\xi$ (see Eqs. \ref{fr} and \ref{SPexp}\,-\,\ref{SPexp1}).
In the calculations, the source radius $R$ for central Au+Au collisions at $\sqrt{s_{NN}}$ = 200 GeV is fixed at 7 fm \cite{Padula06,Zhang15a,Zhy-ijmpe2025},
while the parameter $\xi =$ 4.8 is obtained by fitting the simulated spectra to the experimental data in Fig. \ref{ptr}(a). For other collision energies,
$\xi$ is held constant at 4.8, and $R$ is determined through a similar fitting procedure against the data.
In Table \ref{tab1}, the parameter values used to calculate the transverse momentum spectra and the $\bar{p}/p$ in Fig. \ref{ptr} are shown.
As collision energy decreases, the in-medium proton chemical potential $\mu_p$ increases, while the source radius $R$ and the velocity parameter $v_0$ decrease.
The in-medium mass modification, governed by parameters $\delta m_0$ and $\Lambda$, is implemented with energy-dependent values: $\Lambda$ is set to 1400 MeV at 27, 39, 62.4, and 200 GeV; 1300 MeV at 19.6 GeV; and 1250 MeV at 11.5 GeV. Meanwhile, $\delta m_0$ is fixed at 20 MeV for all collision energies considered.

\begin{table*}
	\caption{\label{tab1}Parameter set for the transverse momentum spectra and $\bar{p}/p$ calculations in Fig. \ref{ptr}.}
	\begin{ruledtabular}
		\begin{tabular}{cccccc}
			Collision energy~&~$\mu_p$ (MeV)&$R$ (fm)\,&~$v_0$&$\delta m_0$(MeV)&$\Lambda$(MeV)\\
            \hline
			200 GeV~&~20~&~7~&~0.7&~20&1400\\
            62.4 GeV~&~65~&~6.17~&~0.63&~20&1400\\
            39 GeV~&~88~&~5.55~&~0.58&~20&1400\\
            27 GeV~&~128~&~5.3~&~0.55&~20&1400\\
            19.6 GeV~&~165~&~5.04~&~0.52&~20&1300\\
            11.5 GeV~&~270~&~4.15~&~0.51&~20&1250\\
		\end{tabular}
	\end{ruledtabular}
\end{table*}

\begin{figure*}[htbp]
\includegraphics[scale=0.69]{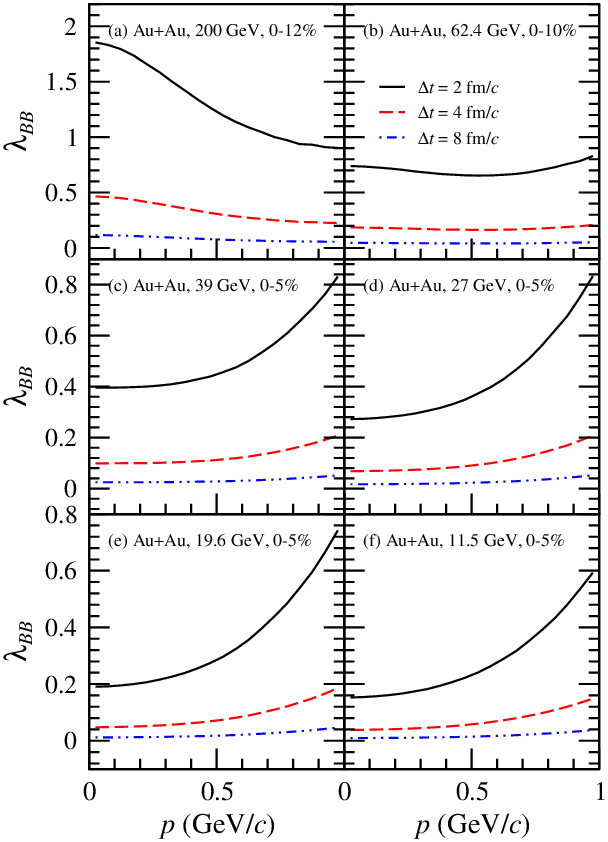}\hspace*{9mm}\includegraphics[scale=0.69]{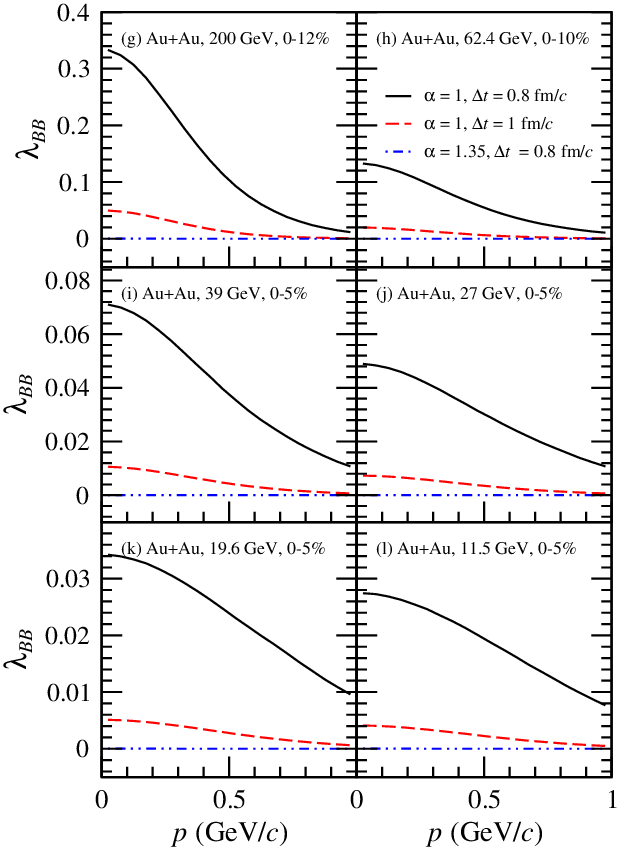}
\vspace*{-2mm}
\caption{(Color online)
Simulated fBBC strengths ($\lambda_{BB}$) of $p\bar{p}$ for central Au+Au collisions at $\sqrt{s_{NN}}$ = 11.5, 19.6, 27, 39, 62.4, and 200 GeV.
The left panel shows the results for sources with a Lorentzian time distribution, while the right panel shows the results for sources with an $\alpha$-stable L\'{e}vy time distribution.}
\label{ckrhic}
\end{figure*}

In Fig. \ref{ckrhic}, the simulated fBBC strengths ($\lambda_{BB}$) of $p\bar{p}$ for central Au+Au collisions at $\sqrt{s_{NN}}$ = 11.5, 19.6, 27, 39, 62.4, and 200 GeV are shown.
The left panel shows the results for sources with a Lorentzian time distribution, while the right panel shows the results for sources $\alpha$-stable L\'{e}vy time distribution.
For each collision system, the in-medium mass modifications of protons and antiprotons and the corresponding source parameters are taken from Table \ref{tab1}.
For sources with a Lorentzian time distribution (Figs. \ref{ckrhic}(a)-(f)), simulations indicate that the fBBC of $p\bar{p}$ yields observable signals unless the time parameter $\Delta t$ is very large ($\Delta t \geq$ 8 fm/$c$). However, experimental data for a momentum around 0.8 GeV/$c$ in central Au+Au collisions at $\sqrt{s_{NN}}$ = 200 GeV show only a $2\%$ signal \cite{BBCsy}, which is inconsistent with the simulations. This discrepancy could be attributed to a large value of $\Delta t$ or the absence of an in-medium mass modification in such a collision.
For central Au+Au collisions at $\sqrt{s_{NN}}$ = 11.5, 19.6, 27, 39, and 62.4 GeV, experimental $p\bar{p}$ fBBC data are currently lacking. Simulations indicate that an observable signal may arise if there is a mass modification, assuming a Lorentzian source time distribution with a sufficiently narrow width.
For sources with an $\alpha$-stable L\'{e}vy time distribution (Figs. \ref{ckrhic}(g)-(l)), at $\sqrt{s_{NN}}$ = 200 GeV, the parameter $\alpha$ is found to be around 1 in the small transverse momentum region; it reaches a value of approximately 1.35 when 0.2$\,<p_T<\,$2 GeV/$c$ \cite{Padula10,Levy-alpha}.
Accordingly, an investigation of the model is conducted with the adoption of these two specific $\alpha$ values. When $\alpha$ = 1.35, the $p\bar{p}$ fBBC signal becomes undetectable due to suppression from the time distribution, leaving almost no observable signal (even when $\Delta t$ = 0.8 fm/$c$).
When $\alpha$ = 1, the fBBC strength for $p\bar{p}$ decreases with increasing momentum. This suggests a potential observable signal in low-momentum regions for central Au+Au collisions at $\sqrt{s_{NN}}$ = 27, 39, 62.4, and 200 GeV, given $\Delta t$ = 0.8 fm/$c$. However, this signal may be suppressed as $\Delta t$ increases.
Moreover, for $\alpha$ = 1 and $\Delta t$ = 0.8 fm/$c$, the simulated fBBC strength at high momenta for central Au+Au collisions at $\sqrt{s_{NN}}$ = 200 GeV closely match the experimental data \cite{BBCsy}.
For central Au+Au collisions at $\sqrt{s_{NN}}$ = 200 GeV, both theory and experiment point toward an $\alpha$-stable L\'{e}vy source time distribution \cite{Levy-alpha}.
Furthermore, fBBC may yield observable signals in the low-momentum region under these conditions.
For central Au+Au collisions at $\sqrt{s_{NN}}$ = 11.5, 19.6, 27, 39, and 62.4 GeV, which currently lack experimental data, model calculations indicate distinct fBBC signatures contingent on the source time distribution: a Lorentzian profile yields strong signals at high momenta, while an $\alpha$-stable L\'{e}vy profile is predicted to produce strong signals at low momenta. Critically, broadening of the time distribution suppresses these signals to unobservable levels in both cases.

\section{Summary and discussion}
In relativistic heavy-ion collisions, particle interactions within the medium induce an in-medium mass modification, which initiates a squeezing effect. This effect leads to the squeezed back-to-back correlation of particle-antiparticle pairs \cite{AsaCsoGyu99,Padula06,Padula10,Padula10a,Zhang15a,YZHANG_CPC15,XuZhang19,fm-PLB,fm-NPA} and also influences single-particle invariant momentum distributions \cite{AsaCsoGyu99,Padula06,fm-PLB,Zhy-ijmpe2025}. This paper studies the impact of the squeezing effect on the transverse momentum spectra of protons and antiprotons. Given that a momentum-independent mass modification fails to describe the experimental spectra and yield ratio \cite{Zhy-ijmpe2025}, a momentum-dependent formulation is used. The range of the in-medium mass modification is constrained by comparing theoretical calculations with experimental data on spectra and yield ratios. {\color{black}The results summarized in Fig. 3 and Table I indicate that STAR experimental data on the single particle spectra of protons and antiprotons, as well as their ratios, are not inconsistent with a momentum-dependent in-medium mass modification of both protons and antiprotons, at six different colliding energies.
These results, together with similar indirect observations of the PHENIX experiment in Ref. \cite{prc2024}, suggest that it is important to search for other observables for indirect observations of in-medium mass-modification of hadrons and the related back-to-back correlations of stable or long-lived particles that do not decay significantly in the hot hadronic matter created in relativistic heavy-ion collisions.}
Based on {\color{black}the mass modification range from Table I}, a theoretical prediction for the $p\bar{p}$ fBBC at RHIC energies is then provided.
Two source time distributions are employed: the Lorentzian distribution and the $\alpha$-stable L\'{e}vy distribution.
In central Au+Au collisions at $\sqrt{s_{NN}}$ = 200 GeV, both theoretical and experimental evidence indicates an $\alpha$-stable L\'{e}vy source time distribution \cite{Levy-alpha}.
Furthermore, under these conditions, fBBC signals may become observable in the low-momentum region.
Model calculations for central Au+Au collisions at $\sqrt{s_{NN}}$ = 11.5, 19.6, 27, 39, and 62.4 GeV (where $p\bar{p}$ fBBC data are absent) show that distinct $p\bar{p}$ fBBC signals emerge depending on the temporal source shape: a Lorentzian profile favors high-momentum signals, while an $\alpha$-stable L\'{e}vy profile favors low-momentum ones. Signal suppression to negligible levels occurs in both scenarios upon broadening of the time distribution.

The effects demonstrated in this paper fundamentally require that the particles experience a change in mass within the medium. In the scenario where no such in-medium modification occurs, the fBBC signal simply does not emerge.
An in-medium mass modification is expected to increase the yield ratio $\bar{p}/p$.
Consequently, within the same collision system, events with a larger $\bar{p}/p$ may present a higher probability for detecting the fBBC, implying a new potential avenue for its observation.
{\color{black}Previous measurements of back-to-back correlations were performed in central collisions \cite{BBCsy}.
 However, the recent PHENIX experiment \cite{prc2024} indicates that the in-medium mass of the ${\eta}'$ meson may be modified even in non-central Au+Au collisions at $\sqrt{s_{NN}}$ = 200 GeV, suggesting that other hadrons may experience similar mass modifications in non-central collisions.
Since the source time distribution is narrower in non-central than in central collisions at the same energy, the suppression of back-to-back correlations is weaker \cite{EPJC2016}.
Thus, in non-central collisions, the back-to-back correlation may also provide observable signals.
Consequently, it is necessary to search for back-to-back correlations not only in central collisions but also in non-central collisions.}

\begin{acknowledgments}
This research was supported by the National Natural Science Foundation
of China under Grant No. 11905085.
\end{acknowledgments}

\end{document}